\documentclass[prl, floatfix, amsmath, twocolumn]{revtex4}

\usepackage{epsfig}
\usepackage{graphicx}            % Include figure files
\usepackage{dcolumn}             % Align table columns on decimal point
\usepackage{bm}                  % bold math
\usepackage{psfrag}

\begin{document}

\title{Quantum Fluctuations in Josephson Junction Comparators}

\author{Thomas~J.~Walls}
\email{twalls@grad.physics.sunysb.edu}
\author{Timur~V.~Filippov}
\author{Konstantin~K.~Likharev}
\affiliation{Department of Physics and Astronomy, Stony Brook University, 
Stony Brook, NY 11974-3800}

\date{\today}

\begin{abstract}
We have developed a method for calculation of quantum fluctuation effects, 
in particular of the uncertainty zone developing at the potential curvature 
sign inversion, for a damped harmonic oscillator with arbitrary time dependence 
of frequency and for arbitrary temperature, within the Caldeira-Leggett model. 
The method has been 
applied to the calculation of the gray zone width $\Delta I_x$ of 
Josephson-junction balanced comparators 
driven by a specially designed low-impedance RSFQ circuit. 
The calculated temperature dependence of $\Delta I_{x}$ in the 
range 1.5 to 4.2K
is in a virtually perfect agreement with experimental data for Nb-trilayer 
comparators with critical current 
densities of 1.0 and 5.5 kA/cm$^2$, without any fitting parameters.
\end{abstract}

\maketitle

The current attention to quantum information processing (see, e.g., the recent 
monograph 
\cite{Nielsen_Chuang}) has renewed interest in fast "single-shot" quantum 
measurements, 
especially in potentially scalable solid-state systems. Among such systems, 
superconductor 
"balanced comparator", based on two similar Josephson junctions (Fig. 1a), 
stands apart as a
very simple, scalable system for which quantum-limited sensitivity has already
been demonstrated experimentally \cite{Filippov_97}.

The device is essentially a SQUID (see, e.g., \cite{Likharev}) in which two 
similar junctions are 
biased in series by a source of Josephson phase difference $\phi_{e} (t)$, 
and in parallel by the current $I_{x}$ to be measured. 
Let the system with $\left| \phi_{e}\right| <\pi$ settle in an equilibrium 
state $\phi = \phi_i$, and then apply a rapid phase 
change $\Delta \phi_e = 2\pi$. (This can be readily done using the so-called 
RSFQ circuitry - see, e.g., the recent review \cite{Bunyk}.) 
As a result, the system becomes statically unstable and the Josephson phase 
$\phi$ has to switch 
to one of adjacent stable states, depending on the sign of $I_x$. 
(For junctions with substantial damping, the 
choice is limited by two states closest to the initial value of 
$\phi$: $\phi_f = \phi_i \pm \pi$).

This process may be readily understood using the "magnetic language": the 
driver circuit providing the pulse $\Delta \phi_e = 2\pi$
in fact injects a single flux quantum into a superconducting loop formed by its 
output stage and the comparator (Fig. 1a). Since the loop is
low-inductive (non-quantizing), the flux quantum has to drop out across one of 
the comparator junctions, depending on the sign of $I_x$. This transient
process produces a large (discrete) output signal, the so-called SFQ pulse 
$V (t)$ with $\int V (t) dt = \Phi_0 = h/2e$ across
the corresponding junction. Such a pulse may be readily picked up and 
registered by relatively crude devices \cite{Bunyk}, 
so that the accuracy of the $I_x$ sign measurement is defined entirely by 
the comparator.
\begin{figure}[t]
\epsfig{file=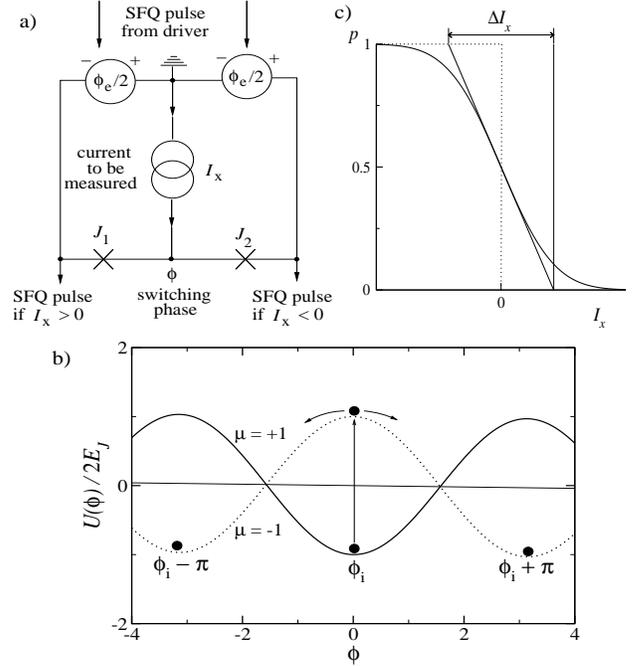, width=3.2in, height=3.7in}
\caption{\label{fig:switch_prob} (a) The balanced comparator, (b) its potential 
energy profile $U(\phi)$
for $I_x = 0.01 I_c$ and two values of the external phase $\phi_e$ 
($\mu = \cos(\phi_e/2)$), and (c) probability $p$ of comparator switching into
state $\phi_f = \phi_i - \pi$ as a function of the measured current $I_x$ 
(schematically).}
\end{figure}

In the absence of fluctuations, the boundary between the two possible outcomes 
would be infinitely sharp - see the dashed line in Fig. 1c; however, 
fluctuations create a finite "gray zone" of $I_{x}$ where probability $p$ 
of switching to a certain finite state changes gradually from 0 to 1 - 
see the solid line in Fig. 1c. The gray zone width, which is traditionally  
defined as
\begin{equation}
\label{eq:smearing_def}
\Delta I_x \equiv \left|\frac{d p}{d I_x}\right|_{I_x = 0}^{-1},
\end{equation}
characterizes the accuracy of the single-shot measurement. This width, 
including its temperature dependence, has been measured in several special 
experiments with externally-shunted Josephson junctions 
\cite{Filippov_95,Filippov_97}. However, theoretically it has been only 
calculated \cite{Filippov_96_2, Filippov_96} for a special function $\phi_e (t)$
 enabling  an analytical solution of the problem, but rather different from 
that used in experiments. Thus, the comparison of theory and 
experiments was not completely conclusive. The goal of this work has been to 
develop a general method of calculation of $\Delta I_x$ for
an arbitrary waveform $\phi_e (t)$ and to compare the results with 
experimental data \cite{Filippov_95,Filippov_97}.

The potential energy of the balanced comparator (Fig. 1a) may be presented in 
the form 
\begin{equation}
\label{eq:pot}
U(\phi) = -2E_J \cos\left[\frac{\phi_e(t)}{2}\right] \cos \phi -
\frac{\hbar}{2e} I_x\phi ,
\end{equation}
where $E_J \equiv \hbar I_c / 2 e$ is the Josephson coupling energy scale, 
and $I_c$ is the critical current of a single junction. 
(In the simplest case, neither $\phi_{e}(t)$ nor $I_x$ depend on the state 
of the comparator.) Notice that the part of "washboard" potential profile, 
contributed by the Josephson junctions, 
changes sign when $\phi_e(t)$ is increased beyond $\pi$. This is exactly the 
reason of the system switching to one of the newly stable states 
$\phi_f \approx \phi_i \pm \pi$  (Fig. 1b). Now let $\left | I_x \right |$ 
and the natural current scales of thermal 
and quantum fluctuations, $I_T = (2e/\hbar)T$ and 
$I_Q = (2e/\hbar)\hbar \omega_p = 2e\omega_p$, respectively 
(where $\omega_p$ is the plasma frequency \cite{Likharev}), all be much 
smaller than $I_c$, and the potential inversion time be of the order of, or 
shorter than the characteristic time of system dynamics. Then the choice of 
the final state is determined by the system 
evolution close to the point $\phi = 0$. In order to describe this evolution, 
we may keep only two leading, linear and quadratic, 
terms in the Taylor expansion of the potential energy (\ref{eq:pot}) near 
this point:
\begin{equation}
\label{eq:lin_pot}
\frac {U(\phi)}{2E_J} = \frac{\mu(t)\phi^2}{2} - \frac{I_x}{2I_c}\phi +
\mbox{const}, \;\; \mu(t) \equiv  \cos \frac{\phi_e(t)}{2}.
\end{equation}
This means that the state choice problem in the original, nonlinear system is 
reduced to that of a damped time-dependent harmonic oscillator with 
frequency $\omega (t)$ defined as $\omega ^2 (t) = \mu (t)\omega_p ^2$, 
where $\mu$ is switched rapidly from a positive initial value $\mu_i$
(in experiments \cite{Filippov_95,Filippov_97}, close to 1) to a negative 
final value $\mu_f = - \mu_i$.

The probability of switching to a final state with $\phi_f < \phi_i$ may be 
found as 
\begin{equation}
\label{eq:p1}
p = \lim_{t \rightarrow \infty} \int\limits_{-\infty}^{\phi_{max}(t)}
\rho(\phi,\phi,t)\,d\phi,
\end{equation}
where $\rho(\phi,\phi^{\prime},t)$ is the system's density matrix traced over 
the degrees of freedom of the environment, and $\phi_{max}(t)$ is 
the coordinate of the
maximum of potential (\ref{eq:lin_pot}) after inversion. Converting to 
coordinates $\eta \equiv \phi + \phi^{\prime}$ and 
$\xi \equiv \phi - \phi^{\prime}$, 
we can express $\rho(\phi,\phi,t) = (1/2) \rho(\eta, 0, t)$ via the system 
propagator $J(\eta, \xi, t | \eta_i, \xi_i,0)$:
\begin{equation}
\label{eq:rho_init}
\rho(\eta, 0, t) = \iint\limits_{-\infty}^{+\infty} J(\eta, 0, t | \eta_i, 
\xi_i,0) \rho (\eta_i,\xi_i,0)\,d\eta_i\,d\xi_i.
\end{equation}

To find the propagator, we may use the Caldeira-Leggett approach 
\cite{Caldeira} with the 
linear distribution of the environment oscillators, 
which gives a quantitatively correct description of systems with externally 
shunted Josephson junctions. According to this theory, 
\begin{gather}
\label{eq:prop1}
J(\eta, \xi, t | \eta_i, \xi_i,0) =  \iint \exp\left[ \frac{i S(\eta,\xi) - 
\theta(\eta,\xi)}{\hbar}\right] D\eta D\xi,
\\
\label{eq:CL_action}
S(\eta,\xi) = \int\limits_0^t \mathcal{L} (\eta,\xi) d\tau - 
\left.\frac{M\gamma}{2}\eta\xi\right|_0^t,
\\
\begin{split}
\theta(\eta,\xi) = &
\frac{2M\gamma}{\pi}\int\limits_0^\Omega d\omega 
\omega\coth(\frac{\hbar\omega}{2T}) 
\\ & \times\int\limits_0^t d\tau \int\limits_0^\tau ds 
\xi(\tau)\xi(s)\cos[\omega(\tau-s)].
\end{split}
\end{gather}
Here, $\mathcal{L} = M (\dot \phi )^2 /2 - M \mu(t) \omega_p ^2 \phi ^2/2$ is 
the Lagrangian of the mechanical oscillator equivalent to our system, 
with mass $M = 2 E_p / \omega_p ^2$, while $\gamma = \omega_p^2/2\omega_c$ is 
its damping parameter,
where $\omega_c \equiv (2e/\hbar)I_c R$, and $R$ is the shunting resistance 
\cite{Likharev}. Parameter $\Omega$ is the cutoff frequency of the 
environment oscillators. To evaluate the path integral (\ref{eq:prop1}), we 
represent coordinates $\eta, \xi$ as  a sum of the 
path parts $\eta (\tau), \xi (\tau)$ minimizing the action $S$, and small 
fluctuations {$\tilde \eta, \tilde \xi$}. The path 
parts satisfy the following equations \cite{Caldeira}:
\begin{eqnarray}
\label{eq:eom1} 
\omega_p^{-2}\frac{d^2\eta}{d \tau^2}  + \omega_c^{-1}\frac{d\eta}{d \tau} + 
\mu(\tau)\eta
= I_x/I_c, \\
\label{eq:eom2}
\omega_p^{-2}\frac{d^2\xi}{d \tau^2}  -  \omega_c^{-1}\frac{d\xi}{d \tau} + 
\mu(\tau)\xi = 0.
\end{eqnarray}

It is convenient to present solutions of these equations as follows: 
\begin{eqnarray}
\label{eq:eom_sol1}
\eta(\tau, t) & = & \eta_i a_1(\tau, t) + \eta a_2(\tau, t) + (I_x / I_c) 
a(\tau, t), \\
\label{eq:eom_sol2}
\xi(\tau, t) & = & \xi_i b_1(\tau, t) + \xi b_2(\tau, t), 
\end{eqnarray}
where functions $a_{1,2}$ and $b_{1,2}$ as functions of $\tau$ obey the
 uniform versions of equations (\ref{eq:eom1}), (\ref{eq:eom2}) with the 
following boundary 
conditions: $a_1(0) = b_1(0) = 1$, $a_1(t) = b_1(t) = 0$, 
$a_2(0) = b_2(0) = 0$, $a_2(t) = b_2(t) = 1$,
while $a$ is the solution to Eq. (\ref{eq:eom1}) with the unit right-hand part 
and boundary conditions $a(0) = a(t) = 0$.

Plugging all these expressions into Eq. (\ref{eq:prop1}), and carrying out a 
lengthy but straightforward (Gaussian) integration, we get
\begin{widetext}
%\begin{equation}
\begin{gather}
\label{eq:J}
J(\eta,\xi,t | \eta_i,\xi_i,0) = F^2(t)\exp\left\{i \left [ K_1\eta_i\xi_i + 
K_2\eta\xi
  - L\eta_i\xi - N\eta\xi_i + \frac{I_x}{I_c} (Q_1\xi_i + Q_2\xi) \right ] - 
[A\xi^2 + B\xi\xi_i +
  C\xi_i^2]\right\},
%\end{equation}
%
%\begin{eqnarray}
\\
\begin{align}
% K1 and K2
\left(\begin{array}{c}K_1 \\ K_2\end{array}\right) & =  
\frac{E_J}{\hbar}\int\limits_0^t d\tau
  \left\{\frac{1}{\omega_p^2}\left(\begin{array}{c}a^{\prime}_1 b^{\prime}_1 \\
        a^{\prime}_2b^{\prime}_2\end{array}\right) - 
\mu(\tau)\left(\begin{array}{c}a_1b_1
        \\ a_2b_2\end{array}\right) +
      \frac{1}{2\omega_c}\left(\begin{array}{c}a_1b^{\prime}_1 - 
a_1^{\prime}b_1 \\
          a_2 b^{\prime}_2 - a_2^{\prime} b_2\end{array}\right) \right\} +
    \frac{E_J}{2\hbar\omega_c}\left(\begin{array}{c}1 \\ 
-1\end{array}\right), \\
% N and L
\left(\begin{array}{c}N \\ L\end{array}\right) & =  
-\frac{E_J}{\hbar}\int\limits_0^t d\tau
  \left\{\frac{1}{\omega_p^2}\left(\begin{array}{c}a^{\prime}_2 b^{\prime}_1 \\
        a^{\prime}_1 b^{\prime}_2\end{array}\right) -
    \mu(\tau)\left(\begin{array}{c}a_2 b_1
        \\ a_1 b_2\end{array}\right) +
      \frac{1}{2\omega_c}\left(\begin{array}{c}a_2 b^{\prime}_1 - 
a_2^{\prime}b_1 \\
          a_1 b^{\prime}_2 - a_1^{\prime} b_2\end{array}\right) \right\}, \\
% Q1 and Q2 
\left(\begin{array}{c}Q_1 \\ Q_2\end{array}\right) & =  
\frac{E_J}{\hbar}\int\limits_0^t d\tau
  \left\{\frac{1}{\omega_p^2}\left(\begin{array}{c}a^{\prime} b^{\prime}_1 \\
        a^{\prime} b^{\prime}_2\end{array}\right) -
    \mu(\tau)\left(\begin{array}{c}a b_1
        \\ a b_2\end{array}\right) +
      \frac{1}{2\omega_c}\left(\begin{array}{c}a b^{\prime}_1 - a^{\prime}b_1 \\
          a b^{\prime}_2 - a^{\prime} b_2\end{array}\right) +
      \left(\begin{array}{c}b_1 \\ b_2\end{array}\right)\right\}, \\
% A, B and C
\left(\begin{array}{c}A \\ B \\ C\end{array}\right) & = 
\frac{E_J}{\pi\hbar\omega_c}\int\limits_0^{\Omega} d\omega
\omega \coth\left(\frac{\hbar\omega}{2T}\right)\int\limits_0^t
\int\limits_0^t d\tau \, d
s\cos[\omega (\tau - s)]\left(\begin{array}{c}b_2(\tau, t)b_2(s, t) \\
    b_1(s, t)b_2(\tau, t) + b_1(\tau, t)b_2(s, t) \\ b_1(\tau, t)b_1(s, t)
\end{array}\right),
%\end{eqnarray}
\end{align}
\end{gather}
\end{widetext}
where $F^{2}(t)$ is a normalization factor, and the prime represents differentiation over $\tau$.

These formulas present the  generalization of Eq. (6.26) of Ref. 
\cite{Caldeira} to the case of arbitrary time dependence of the 
oscillator potential curviture $\mu (t)$. Equations (\ref{eq:rho_init}), 
(\ref{eq:J}) show that if the initial density matrix
is Gaussian (as it is, e.g., for a system in thermal equilibrium), the final 
matrix is also Gaussian, 
with the average phase $\langle \phi \rangle$ and variance 
$\langle \phi^2 \rangle$ determined by parameters $K_1, N, Q_1$ and $C$. 
(Other parameters affect only the final phase velocity distribution, which is 
not important for our particular problem.)

Using the definition (\ref{eq:smearing_def}), the gray zone width may now be 
calculated as
\begin{equation}
\begin{split}
\label{eq:ix_gen}
\Delta I_x = & \lim_{t \rightarrow \infty} \frac{\left( 2\pi 
\left\langle \phi^{2}\right\rangle \right) ^{1/2}}{\left| \frac{d}{dI_{x}}
\left\langle \phi \right\rangle \right| } \\
= & 2\pi^{1/2} I_c 
\frac{\left[ C+4K_{1}^{2}\langle \phi^{2} \rangle_{i} 
+ \left( I_{c}/2e\omega _{c}\right) ^{2}\langle \dot{\phi}^{2}\rangle _{i}
\right] ^{1/2}}{K_{1}\mu_i^{-1}+Q_{1}}.
\end{split}
\end{equation}
This is our central result. For the particular case of the RSFQ driver circuit 
used in 
experiments \cite{Filippov_95,Filippov_97}, the 
function $\mu (t)$ has been calculated numerically from the circuit schematics, 
using the 
PSCAN software package \cite{Polonsky}. 
Since functions $a(\tau, t)$ and $b(\tau, t)$ are exponential near the boundary 
points $\tau = 0$ and $\tau = t$, 
the standard "shooting" methods for the numerical calculation of these 
functions would be unstable. 
Because of this we have used the relaxation 
method \cite{Keller}. Upon the calculation of $a(\tau, t)$ and $b(\tau, t)$, 
parameters $K_1$ and $Q_1$ were obtained by the standard numerical 
integration using the trapezoidal approximation \cite{Press}. $C$ was 
calculated using 3D Monte Carlo integration
where an integration by parts helps control the discontinuity at 
$\omega \rightarrow 0$.  
Due to the shape of the 
function $b_1$, a combination of stratified and importance sampling greatly 
increases the convergence 
time, so a variant of the VEGAS algorithm \cite{Press}
was used. The bath oscillator cutoff
frequency $\Omega$ was taken large enough $(50\omega_p)$ to avoid any effect 
on the calculation results.

\begin{figure}[t]
\epsfig{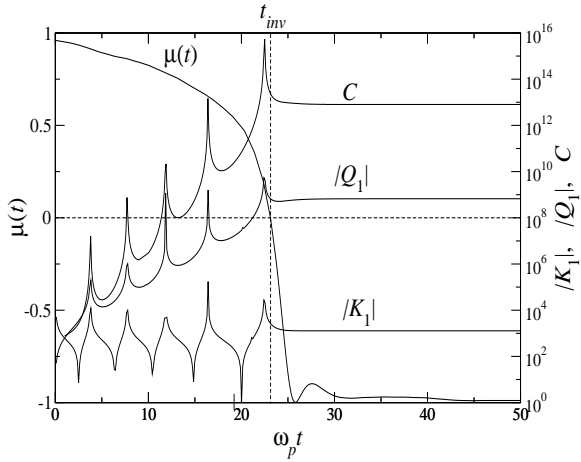}
\caption{\label{fig:muab} Calculated function $\mu (t)$ and parameters of the 
final phase distribution for the RSFQ drivers used in experiments 
\cite{Filippov_95,Filippov_97} for $\beta_c = 1$.  The time scale 
$\omega_p^{-1}$ is 
close to 1.1 ps for the critical current density  $j_c$ = 1 kA/cm$^2$ 
\cite{Filippov_95} and 0.47 ps for $j_c$ = 5.5 kA/cm$^2$ \cite{Filippov_97}.}
\end{figure}

\begin{figure}[t]
\psfrag{h}{$\hbar$}
\epsfig{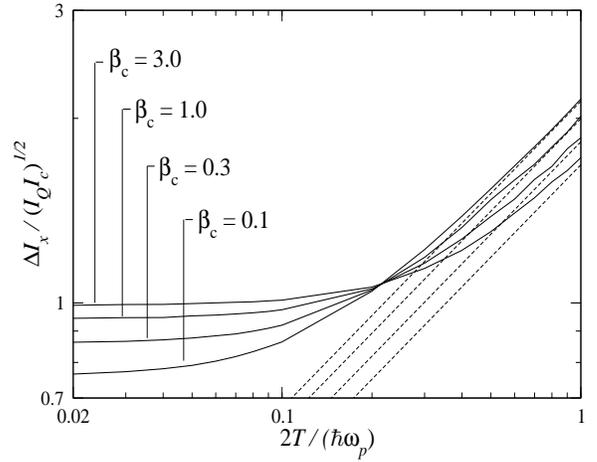}
\caption{\label{fig:pscan_betas} Temperature dependence of $\Delta I_x$ for
  $\mu (t)$ shown in Fig. 2, calculated for several values of $\beta_c$. 
Dashed lines represent the thermal limit.}
\end{figure}

Figure  \ref{fig:muab} shows $\mu$ and the essential parameters of the 
Gaussian distribution as functions of $t$. 
One can see that if the interval 
$[0,t]$ includes the time point $t_{inv}$ of the potential curvature sign 
inversion, with both $t_{inv}$ and 
$t - t_{inv}$ much longer than the oscillator's reciprocal 
bandwidth 
$\Delta \omega^{-1} \approx \max [1/ \omega_{c}, \omega_c/2\omega_{p}^{2}]$, 
then $C \gg 1$, and $Q_1 \gg K_1$, so that the final density matrix and 
switching probability $p$ 
do not depend on the initial state of the system. 
In this limit, Eq. (\ref{eq:ix_gen}) takes a very simple form:
\begin{equation}
\label{eq:ix}
\Delta I_x = 2\pi^{1/2}I_c \frac{C^{1/2}}{Q_1}.
\end{equation}
Figure \ref{fig:pscan_betas} shows the resulting temperature dependence of the 
gray zone width for several values of the inertia parameter 
(normalized junction capacitance) $\beta_c \equiv (\omega_c/\omega_p)^2$. At 
high temperatures, $\Delta I_x$ grows as $T^{1/2}$ due to thermal fluctuations,
while at $T \rightarrow 0$ it saturates due to quantum fluctuations. Note also 
that the dependence of $\Delta I_x$ on $\beta_c$ is
different for high and low temperatures: if thermal fluctuations dominate, the 
gray zone width depends on $\beta_c$ only weakly, saturating at 
comparable values at both $\beta_c \rightarrow 0$ and 
$\beta_c \rightarrow \infty$. However, in the quantum fluctuation range 
($T \rightarrow 0$), 
$\Delta I_x$ grows as $\beta_c^{1/4}$ at high damping 
($\beta_c \rightarrow 0$) and saturates in the opposite limit of low damping.
All these dependences may be qualitatively understood from the following 
simple consideration: $\Delta I_x$ crudely 
equals to the signal current that creates the phase shift $\phi = I_x/2I_c$ 
equal to the r.m.s. value of phase noise in thermal equilibrium. 
The latter value may be estimated assuming that an equivalent current noise 
source \cite{Likharev} with equilibrium spectral density $S_{I}(\omega )= (4/R) 
(\hbar \omega/2 )\coth (\hbar \omega/2T)$ acts on a time-independent linear 
oscillator within the bandwidth $\Delta \omega$ defined above.

\begin{figure}[t]
\epsfig{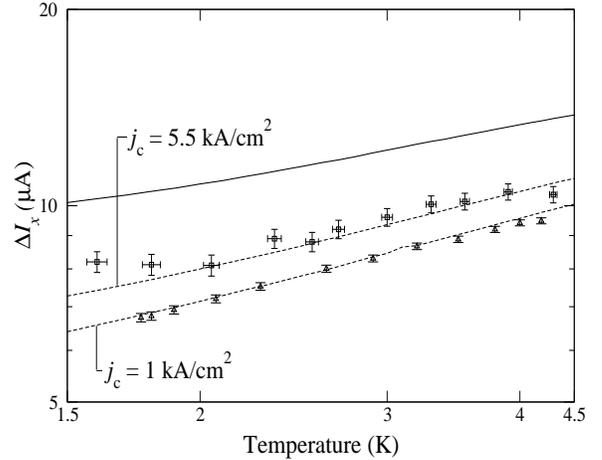}
\caption{\label{fig:smearing} Temperature dependence of $\Delta I_x$. Points 
show  experimental data from Refs. \cite{Filippov_95,Filippov_97}. 
The dashed lines are results of calculation taking into account 
the Ambegoakar-Baratoff temperature dependence of $I_c$. The solid lime is 
the theory for an instantaneous change of $\mu$ from +1 to -1.}
\end{figure}

Figure \ref{fig:smearing} shows the comparison of our results with 
experimental data for comparators based on niobium-trilayer (Nb/AlO$_x$/Nb) 
Josephson 
junctions with $I_c \mid_{T = 4.2 K}$ = 145 $\mu A$, $\beta_c = 1$, for two 
values of the critical current density: $j_c$ = 1 kA/cm$^2$ \cite{Filippov_95} 
and 5.5 kA/cm$^2$ \cite{Filippov_97}. One can see that besides the deviation 
of the two lowest-$T$ points in experiments \cite{Filippov_95}, which was 
apparently caused by sample self-heating, the theory gives a virtually 
perfect description of experimental results, without 
any fitting parameters. (The possibility of a substantial external noise 
contribution to $\Delta I_x$ in experiments 
\cite{Filippov_95,Filippov_97} has been ruled out by special control 
experiments using similar comparators, fabricated on the same chip, 
but driven by "softer" waveforms.)

To summarize, we have developed a method of analysis of quantum fluctuations 
at the inversion of the potential curvature sign of a damped harmonic 
oscillator. When applied to Josephson junction comparator, these results may 
be used for numerical caculation of the gray zone width 
$\Delta I_x$. Such calculation for the Nb-trilayer comparators 
\cite{Filippov_95, Filippov_97} gave a nearly perfect agreement with 
experimental data.
Our result may be also generalized to the case of a finite inductive 
impedance of the source of the signal $I_x$, which is typical for 
Josephson junction systems, e.g., magnetic flux qubits \cite{Lukens,Mooij}. 
Indeed, the impedance may be described by connecting the source 
inductance $L$ in parallel with the source of $I_x$ in Fig. 1a. An elementary 
calculation shows that this leads to re-normalization of
function $\mu (t)$:
\begin{equation}
\mu (t) \rightarrow \cos \left[ \frac{\phi _{e}(t)}{2}\right] +
\frac{1}{2\lambda },\text{
\ \ }\lambda \equiv 2\pi \frac{LI_{c}}{\Phi _{0}}.
\end{equation}
This means that if the inductance is not too low, $\lambda > 1/2$, the input 
SFQ pulse $\Delta \phi_e = 2 \pi$ develops an instability of phase $\phi$ 
just as was described above, 
and our theory gives a ready recipe for the calculation of $\Delta I_x$ and 
hence of the signal flux resolution $\Delta \Phi_x = L \Delta I_x$. 

However, in order to reduce dephasing, flux
qubits typically require unshunted Josephson junctions. This is why a natural 
next task would be a calculation of $\Delta \Phi_x$ for the case when
damping is dominated by quasiparticle tunneling in unshunted junctions. For 
this, the Caldeira-Legget action (\ref{eq:CL_action}) 
should be replaced 
with one found by Ambegaokar $\it {et}$ $\it {al.}$  \cite{AES}.
  
Useful discussions with D. V. Averin, J. E. Lukens, Yu. A. Polyakov, and V. K. 
Semenov are gratefully acknowledged. The work was supported in part by 
DoD, ARDA, and AFOSR as a part of DURINT program.

\bibliography{prl02_12}

\begin{thebibliography}{14}
\expandafter\ifx\csname natexlab\endcsname\relax\def\natexlab#1{#1}\fi
\expandafter\ifx\csname bibnamefont\endcsname\relax
  \def\bibnamefont#1{#1}\fi
\expandafter\ifx\csname bibfnamefont\endcsname\relax
  \def\bibfnamefont#1{#1}\fi
\expandafter\ifx\csname citenamefont\endcsname\relax
  \def\citenamefont#1{#1}\fi
\expandafter\ifx\csname url\endcsname\relax
  \def\url#1{\texttt{#1}}\fi
\expandafter\ifx\csname urlprefix\endcsname\relax\def\urlprefix{URL }\fi
\providecommand{\bibinfo}[2]{#2}
\providecommand{\eprint}[2][]{\url{#2}}

\bibitem[{\citenamefont{Nielsen and Chuang}(2000)}]{Nielsen_Chuang}
\bibinfo{author}{\bibfnamefont{M.~A.} \bibnamefont{Nielsen}} \bibnamefont{and}
  \bibinfo{author}{\bibfnamefont{I.~L.} \bibnamefont{Chuang}},
  \emph{\bibinfo{title}{Quantum Computation and Quantum Information}}
  (\bibinfo{publisher}{Cambridge U. Press, Cambridge, UK},
  \bibinfo{year}{2000}).

\bibitem[{\citenamefont{Semenov et~al.}(1997)}]{Filippov_97}
\bibinfo{author}{\bibfnamefont{V.~K.} \bibnamefont{Semenov}}
  \bibnamefont{et~al.}, \bibinfo{journal}{IEEE Trans. \ on Appl. \ Supercond.}
  \textbf{\bibinfo{volume}{7}}, \bibinfo{pages}{3617} (\bibinfo{year}{1997}).

\bibitem[{\citenamefont{Likharev}(1986)}]{Likharev}
\bibinfo{author}{\bibfnamefont{K.~K.} \bibnamefont{Likharev}},
  \emph{\bibinfo{title}{Dynamics of Josephson Junctions and Circuits}}
  (\bibinfo{publisher}{Gordon and Breach, New York}, \bibinfo{year}{1986}).

\bibitem[{\citenamefont{Bunyk et~al.}(2000)\citenamefont{Bunyk, Likharev, and
  Zinoviev}}]{Bunyk}
\bibinfo{author}{\bibfnamefont{P.}~\bibnamefont{Bunyk}},
  \bibinfo{author}{\bibfnamefont{K.}~\bibnamefont{Likharev}}, \bibnamefont{and}
  \bibinfo{author}{\bibfnamefont{D.}~\bibnamefont{Zinoviev}},
  \bibinfo{journal}{Int. J. of High Speed Electron. and Syst.}
  \textbf{\bibinfo{volume}{11}}, \bibinfo{pages}{257} (\bibinfo{year}{2000}).

\bibitem[{\citenamefont{Filippov et~al.}(1995)}]{Filippov_95}
\bibinfo{author}{\bibfnamefont{T.~V.} \bibnamefont{Filippov}}
  \bibnamefont{et~al.}, \bibinfo{journal}{IEEE Trans. \ on Appl. \ Supercond.}
  \textbf{\bibinfo{volume}{5}}, \bibinfo{pages}{2240} (\bibinfo{year}{1995}).

\bibitem[{\citenamefont{Filippov}(1995)}]{Filippov_96_2}
\bibinfo{author}{\bibfnamefont{T.~V.} \bibnamefont{Filippov}},
  \bibinfo{journal}{JETP Letters} \textbf{\bibinfo{volume}{61}},
  \bibinfo{pages}{858} (\bibinfo{year}{1995}).

\bibitem[{\citenamefont{Filippov}(1996)}]{Filippov_96}
\bibinfo{author}{\bibfnamefont{T.~V.} \bibnamefont{Filippov}},
  \bibinfo{journal}{Russian Microelectronics} \textbf{\bibinfo{volume}{25}},
  \bibinfo{pages}{250} (\bibinfo{year}{1996}).

\bibitem[{\citenamefont{Caldeira and Leggett}(1983)}]{Caldeira}
\bibinfo{author}{\bibfnamefont{A.~O.} \bibnamefont{Caldeira}} \bibnamefont{and}
  \bibinfo{author}{\bibfnamefont{A.~J.} \bibnamefont{Leggett}},
  \bibinfo{journal}{Physica} \textbf{\bibinfo{volume}{121 A}},
  \bibinfo{pages}{587} (\bibinfo{year}{1983}).

\bibitem[{\citenamefont{Polonsky et~al.}(1991)\citenamefont{Polonsky, Semenov,
  and Shevchenko}}]{Polonsky}
\bibinfo{author}{\bibfnamefont{S.~V.} \bibnamefont{Polonsky}},
  \bibinfo{author}{\bibfnamefont{V.~K.} \bibnamefont{Semenov}},
  \bibnamefont{and} \bibinfo{author}{\bibfnamefont{P.~N.}
  \bibnamefont{Shevchenko}}, \bibinfo{journal}{Spercond.\ Sci. \ Technol.}
  \textbf{\bibinfo{volume}{4}}, \bibinfo{pages}{667} (\bibinfo{year}{1991}).

\bibitem[{\citenamefont{Keller}(1992)}]{Keller}
\bibinfo{author}{\bibfnamefont{H.~B.} \bibnamefont{Keller}},
  \emph{\bibinfo{title}{Numerical Methods for Two-Point Boundary-Value
  Problems}} (\bibinfo{publisher}{Dover}, \bibinfo{year}{1992}).

\bibitem[{\citenamefont{Press et~al.}(1997)}]{Press}
\bibinfo{author}{\bibfnamefont{W.~H.} \bibnamefont{Press}}
  \bibnamefont{et~al.}, \emph{\bibinfo{title}{Numerical Recipes in C}}
  (\bibinfo{publisher}{Cambridge}, \bibinfo{year}{1997}).

\bibitem[{\citenamefont{Friedman et~al.}(2000)}]{Lukens}
\bibinfo{author}{\bibfnamefont{J.~R.} \bibnamefont{Friedman}}
  \bibnamefont{et~al.}, \bibinfo{journal}{Nature}
  \textbf{\bibinfo{volume}{406}}, \bibinfo{pages}{43} (\bibinfo{year}{2000}).

\bibitem[{\citenamefont{van~der Wal et~al.}(2000)}]{Mooij}
\bibinfo{author}{\bibfnamefont{C.~H.} \bibnamefont{van~der Wal}}
  \bibnamefont{et~al.}, \bibinfo{journal}{Science}
  \textbf{\bibinfo{volume}{290}}, \bibinfo{pages}{773} (\bibinfo{year}{2000}).

\bibitem[{\citenamefont{Ambegaokar et~al.}(1982)\citenamefont{Ambegaokar,
  Eckert, and Sch{\"{o}}n}}]{AES}
\bibinfo{author}{\bibfnamefont{V.}~\bibnamefont{Ambegaokar}},
  \bibinfo{author}{\bibfnamefont{U.}~\bibnamefont{Eckert}}, \bibnamefont{and}
  \bibinfo{author}{\bibfnamefont{G.}~\bibnamefont{Sch{\"{o}}n}},
  \bibinfo{journal}{Phys. Rev. Lett.} \textbf{\bibinfo{volume}{48}},
  \bibinfo{pages}{1745} (\bibinfo{year}{1982}).

\end{thebibliography}

\end{document}